\documentclass[twocolumn,aps,prx,nofootinbib,floatfix,longbibliography,groupedaddress]{revtex4-2}
\usepackage{amsmath,bbm}
\usepackage{graphicx}
\usepackage{multirow}
\usepackage{hyperref}
\usepackage{amsfonts}
\usepackage{amsbsy}
\usepackage{xcolor}
\usepackage{soul}
\usepackage{graphicx}
\usepackage{mathtools}
\usepackage{bm}

\renewcommand{\vec}[1]{\mathbf{#1}}
\newcommand{\vecg}[1]{\bm{#1}}

\newcommand{\idop}{\mathbbm{1}}

\newcommand{\Pc}{\mathcal{P}}
\newcommand{\cbE}{\boldsymbol{\mathbf{\cal E}}}
\newcommand{\bra}[1]{\ensuremath{\left\langle #1 \right\vert}}
\newcommand{\ket}[1]{\ensuremath{\left\vert #1 \right\rangle}}
\newcommand{\unitvec}[1]{\hat{\mathbf{{#1}}}}
\renewcommand{\vec}[1]{\mathbf{#1}}

\begin{document}
\title{Unidirectional absorption, storage, and emission of single photons in a collectively responding bilayer atomic array}

\date{\today}
\author{K.~E.~Ballantine}
\email{k.ballantine@lancaster.ac.uk}
\author{J.~Ruostekoski}
\email{j.ruostekoski@lancaster.ac.uk}

\affiliation{Department of Physics, Lancaster University, Lancaster, LA1 4YB, United Kingdom}

\begin{abstract}

Two-dimensional regular arrays of atoms are a promising platform for quantum networks, with collective subradiant states providing long-lived storage and collimated emission allowing for natural coherent links between arrays in free space. However, a single-layer lattice can only efficiently absorb or emit light symmetrically in the forward and backward directions. Here we show how a bilayer lattice can absorb a single photon either incident from a single direction or an arbitrary superposition of forward and backward propagating components. The excitation can be stored in a subradiant state, transferred coherently between different subradiant states, and released, again in an arbitrary combination of highly collimated forward and backward propagating components. 
We explain the directionality of single and bilayer arrays by a symmetry analysis based on the scattering parities of different multipole radiation components of collective excitations. The collective modes may exhibit the conventional half-wave loss of fields
near the array interface or completely eliminate it.
The proposed directional control of absorption and emission paves the way for effective one-dimensional quantum communication between multiple arrays, with single-photons propagating backward and forward between quantum information-processing and storage stages. 

\end{abstract}

\maketitle

\section{Introduction}

Two-dimensional (2D) atomic arrays interacting with light~\cite{Jenkins2012a,Bettles2015d,Bettles2016,Facchinetti16,Yoo2016,Kramer2016,Shahmoon,
Bettles2017,Perczel2017b,Plankensteiner2017,Jen17,Asenjo-Garcia2017a,Mkhitaryan18,Yoo_2018,Orioli19,Javanainen19,
Williamson2020b,Cidrim20,Parmee2020,Parmee20bistable,Alaee20,Ballantine20Huygens,Bekenstein2020,
Yoo20,Bettles20,Ballantine21PT,Zhang22,Cardoner21,Ballantine21quantum,Rubies21,Cosimo2021}  provide a novel quantum-mechanical many-body system to exploit collective effects in the control of light. 
Emission from large lattices is highly collimated~\cite{Grankin18,Facchinetti18,Guimond2019,Javanainen19,Ballantine20ant}, leading to efficient coherent quantum links in free space between them. 
Quantum networks~\cite{ChoiEtAlNature2008,Ritter12} require just such coherent links between nodes, while the effective 1D scattering also eliminates spontaneous emission in different directions, which is a major loss channel for single atoms or random atomic ensembles~\cite{HAM10}. 
Experimental control of 2D arrays of cold atoms is also rapidly improving~\cite{Bakr10,ShersonEtAlNature2010,Lester15,Kim16,
Barredo18,Cooper18,Saskin19,Mello19,Rui2020}, and the collective narrowing of the reflection linewidth beyond the single-atom limit has been recently observed~\cite{Rui2020}, analogously to the steady-state driving of giant subradiant states of 2D meta-atom arrays~\cite{Jenkins17}. Further progress towards truly quantum manipulation, however, requires directional control over the efficient absorption, storage, and release of propagating single photons.

Subradiant states in free-space 2D arrays of atoms~\cite{Jenkins2012a,Bettles2015d,Facchinetti16,Bettles2016,Asenjo-Garcia2017a,Shahmoon,Perczel2017b,Jen17,
Guimond2019,Parmee2020,Williamson2020b,Bettles20,Ballantine20ant} are promising for light storage due to their isolation from the environment, 
but, unlike in the case of 1D chains of atoms~\cite{Williamson2020,Holzinger21}, in 2D arrays strongly subradiant modes cannot be directly driven by incident fields. Previous proposals have demonstrated how such modes can be excited in the steady state, using atomic level shifts to control the orientation of the dipoles~\cite{Facchinetti16,Facchinetti18}, or by optimizing driving~\cite{Manzoni18}. Besides the giant experimental subradiance of regular subwavelength arrays~\cite{Rui2020, Jenkins17}, long-lived states have also been experimentally observed in disordered atomic clouds~\cite{Guerin_subr16,Ferioli21}. More recently, schemes to achieve efficient absorption of a time-dependent single-photon pulse have been put forward~\cite{Ballantine21quantum,Rubies21}.
However, for a single thin layer, complete absorption is only possible when illuminated symmetrically from both sides. This is because, during the absorption, it is necessary to first excite the atomic dipoles, which then radiate during the process, leading to scattering losses. Such scattering can be suppressed by interference with the incident beam, but only for transmitted light, while reflected light is lost.

Here we show how interference of scattering from a bilayer lattice can be controlled, depending on symmetry, in order to absorb and emit light in the forward direction, backward direction, or an arbitrary combination of the two. We show how a photon incident from a single direction can be absorbed with probability $0.79$, rising to $0.93$ when the time-dependence of the incident pulse is optimized, well in excess of the theoretical limit of $0.5$ for a single layer. The storage occurs by first exciting a superposition of in-plane excitations, the polarization amplitudes of which are either symmetric or antisymmetric between the two atomic layers, and then rotating the polarizations via a linear Zeeman splitting. Each in-plane mode is thus coupled to a subradiant, out-of-plane mode with corresponding symmetry. We further demonstrate how coupling between the out-of-plane modes can be used to control the storage, leading to coherent oscillations between the subradiant modes, and how, by engineering the correct phase between these modes, the excitation can be transferred back to the in-plane modes in such a way that it is emitted in the forward direction, backward direction, or an arbitrary combination of the two, independent of the incident beam direction. 
The proposed scheme for highly directional single photon storage and emission is also very different from the protocols of single photon sources based on the atomic
ensembles in the Rydberg states~\cite{Petrosyan18}. In such a scheme, a single photon is transferred to an atomic ensemble via Rydberg excitations using, e.g.,  STIRAP, while assuming
all the atoms being confined inside the Rydberg dipole blockade region. 

The directionality of light coupling is explained in terms of the odd and even parity of the scattered light from the collective excitation eigenmodes of the arrays that depends on their electric and magnetic multipole character. This is another example of symmetry properties of
scattering in planar atomic arrays that also plays a role in producing Huygens' surfaces~\cite{Ballantine20Huygens} and in generating a ${\cal PT}$ scattering symmetry~\cite{Ballantine21PT}.
We show that more complex geometries, with more than two atoms per unit cell, can also be used to engineer pure magnetic dipole or electric quadrupole modes with the necessary symmetry.

\section{Light-matter relations}

\begin{figure}[t]
 \centering
  \includegraphics[width=\columnwidth]{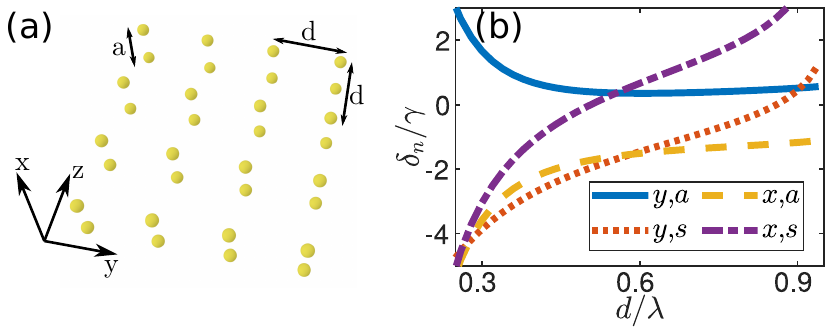}
 \caption{(a) Geometry of atomic bilayer lattice in $yz$ plane with lattice constant $d$ and layer separation $a$, capable of absorbing and emitting unidirectional light. (b) Collective excitation eigenmode level shifts $\delta_n$ of the mode with dipoles oscillating in-plane and symmetric between top and bottom layers $\Pc_{ys}$, the in-plane antisymmetric mode $\Pc_{ya}$, out-of-plane symmetric mode $\Pc_{xs}$, and out-of-plane antisymmetric mode $\Pc_{xa}$. Level shifts of $\Pc_{ya}$ and $\Pc_{ys}$ intersect at $d=0.91\lambda$. $20\times 20\times 2$ lattice with $a=0.25\lambda$. 
  }
 \label{fig:fig1}
\end{figure}

We consider a square lattice in the $yz$ plane with lattice constant $d$. While in Sec.~\ref{sec:mono} we briefly recap the case of a single layer, with one atom per unit cell, and in Sec.~\ref{sec:hs} we consider a four-atom unit cell, the majority of this work focuses on the case of two atoms per unit cell forming a simple bilayer lattice, each atom of the unit cell located in the different layer, as shown in Fig.~\ref{fig:fig1}. The layer separation is $a$, i.e.\ atomic positions $x = \pm a/2$, $y=md$, $z=nd$ for integer $m,n$, as could be achieved by an additional double-well structure of the trap in the $x$ 
direction~\cite{Koepsell20,Gall21} or using optical tweezers~\cite{Barredo18}. There is one atom per site and a $J=0\rightarrow J^{\prime}=1$ transition. Systems with flexible parameter values could be achieved, e.g., in Sr atom arrays~\cite{Olmos13}.
 A single-photon pulse is incident in the $\pm \unitvec{x}$ direction with $y$ polarization. Since the quantization axis is perpendicular to the propagation axis and the polarization, the incident field drives a combination of the transitions to the $J^\prime=1,m=\pm 1$ levels, while the $m=0$ level remains unoccupied. 

Here we consider the interaction of the lattice with a single photon, and so limit the atomic dynamics to at most one excitation\footnote{The analogies and differences between the dynamics of single-excitation and that of the coupled-dipole model of classical harmonic oscillators in the limit of low light intensity are discussed in detail in Ref.~\cite{Ballantine21quantum}.}. 
The basic formalism is based on nonrelativistic electrodynamics~\cite{CohenT,PowerBook}.
The reduced many-body density matrix for the atoms can thus be split into one-excitation and zero-excitation parts, $\rho=\ket{\Psi}\bra{\Psi}+p_G\ket{G}\bra{G}$, with $p_G$ the probability that the lattice is in the ground state $\ket{G}=\prod_j\ket{g_j}$.
The single excitation sector is described by the state $\ket{\Psi}=\sum_{\mu j} \Pc_{\mu}^{(j)}\hat{\sigma}_{\mu j}^{+}\ket{G}$ with $\hat{\sigma}_{\mu j}^{+}=\ket{e_{\mu j}}\bra{g_j}$ the raising operator to level $\mu$ on atom $j$~\cite{Ballantine20ant}. (Here, and in the following, the atomic amplitudes and the light fields refer to the slowly varying positive-frequency components, with rapid fluctuations $\sim\exp{(-i\Omega t)}$, at the frequency $\Omega=ck$ of the incident electric field, filtered out.)    
The dynamics can be written in matrix form, with $\vec{b}$ containing the amplitude components $\Pc_\mu^{(j)}$ for $j=1,\ldots,N$ and $\mu=0,\pm 1$ which evolve according to~\cite{Ballantine20ant,Lee16}
\begin{equation}
\label{eq:eom}
\dot{\vec{b}}=i(\mathcal{H}+\delta\mathcal{H})\vec{b}+ \vecg{\zeta}.
\end{equation} 
The $3N\times 3N$ non-Hermitian matrix $\mathcal{H}$ describes interactions between atoms, with off diagonal components
\begin{equation}
i\xi \unitvec{e}_\mu^{\ast}\cdot \mathsf{G}(\vec{r}_j-\vec{r}_k)\unitvec{e}_\nu,
\end{equation} where $\xi=6\pi\gamma/k^3$ with $\gamma=\mathcal{D}^2k^3/(6\pi\hbar\epsilon_0)$ the single-atom linewidth and $\mathcal{D}$ the reduced dipole matrix element, and $\mathsf{G}$ is the standard dipole radiation kernel
such that the scattered field $\epsilon_0\vec{E}(\vec{r})=\mathsf{G}(\vec{r})\vec{d}$ at a point $\vec{r}$ from a dipole $\vec{d}$ at the origin is 
\begin{multline}\label{eq:rad_kernel_def}
\mathsf{G}(\vec{r})\vec{d} = -\frac{\vec{d}\delta(\vec{r})}{3}+\frac{k^3 e^{ikr}}{4\pi}\left\{(\unitvec{r}\times\unitvec{d})\times\unitvec{r}\frac{1}{kr}
\right. \\
\left. -\left[ 3\unitvec{r}(\unitvec{r}\cdot\unitvec{d})-\unitvec{d} \right]\left[ \frac{i}{(kr)^2}-\frac{1}{(kr)^3} \right]
 \right\},
\end{multline} 
with $r=|\vec{r}|$ and $\hat{r}=\vec{r}/r$~\cite{Jackson}. Note the contact term is inconsequential to the physics for ideal dipoles~\cite{Ruostekoski1997b,Lee16}, and vanishes for hard-core bosons which cannot overlap. 
The diagonal entries of $\mathcal{H}$ are $\delta\Omega+i\gamma$, where $\delta\Omega=\Omega-\omega$ is the detuning of the incident field from the single-atom resonance $\omega$. The diagonal $\delta{\cal H}$ contains additional level shifts $\Delta_{\mu}^{(j)}=\omega-\omega^{(j)}_{\mu}$. The recoil of the atoms~\cite{Robicheaux19} is assumed to be negligible.
The single-photon pulse is approximated by a time-dependent driving term $\vecg{\zeta}$, with components
\begin{equation}
i(\xi/\mathcal{D})\unitvec{e}_\mu^\ast\cdot\epsilon_0\cbE(\vec{r}_j,t)
\end{equation}
driving level $\mu$ of atom $j$.
The incident field is
 \begin{equation}
\label{eq:drive}
\cbE(\vec{r},t)=\mathcal{N} f(x) g(\vec{r}_j) h(t) \unitvec{e}_{y},
\end{equation} 
with $g(\vec{r})$ the Gaussian spatial amplitude, $h(t)$ a time-dependent envelope which varies on timescales much slower than that set by the optical frequency, $\mathcal{N}$ a normalization factor chosen such that the total energy of the beam is $\hbar\Omega$, and an overall propagation phase 
\begin{align}
\label{eq:direction}
f(x) &= \alpha\exp{(ikx)}+\beta\exp{(-ikx)},\\
&= (\alpha+\beta)\cos{(kx)}+i(\alpha-\beta) \sin{(kx)},
\end{align} 
determining the incident direction and symmetry, such that $|\alpha|^2+|\beta|^2=1$. Hence, $\alpha$ is the amplitude of light propagating in the $+\unitvec{x}$ forward direction over time, and $\beta$ that in the $-\unitvec{x}$ back direction\footnote{We can choose the lattice to be centered at $x=0$ without loss of generality, since a shift in the lattice is equivalent to a change in phase of $\alpha$ and $\beta$.}. The analysis of a time-dependent pulse is a key ingredient in understanding the response to a finite energy single-photon drive. While the classical response of the lattice to coherent light in the limit of low light intensity would obey similar dynamics, the resulting state would be very different, with only the single-photon excitation case resulting in quantum entanglement~\cite{Ballantine21quantum}.

\section{Absorption in single layer}
\label{sec:mono}

For a 2D layer confined to a single plane, the maximum efficiency of absorption of a pulse of light traveling in a single direction is $0.5$. Taking $\alpha=1$, for example, gives $f(x)=\exp{(ikx)}=\cos{(kx)}+i\sin{(kx)}$ in Eq.~(\ref{eq:direction}). Then, for a lattice at $x=0$, only the symmetric $\cos{(kx)}$ component drives the atomic dipoles, while the remaining part of the incident field with 50\% of intensity is irrelevant. Indeed, the electric field at $x=0$ is exactly the same as for a standing wave, $\alpha=\beta=1/2$, but this configuration has only half the total incident energy as the unidirectional beam. 

This same limitation can be understood in a different light by considering that any absorption event, including any possible temporal or spatial variation of level shifts, has a time-reversed equivalent which corresponds to an allowed process of the excited lattice emitting the light. For a single layer, regardless of the state the excitation occupies or the level shifts, there is no way to break the symmetry between forward and backward, and so all emission processes have equal total power radiated in both these directions. Hence, the reverse absorption process will result in scattering losses when the lattice is not illuminated symmetrically. When light is incident symmetrically from both sides, it can indeed be absorbed with near-unity efficiency~\cite{Roger15} which can be utilized in storing a single photon~\cite{Ballantine21quantum,Rubies21}. For multiple layers, as described further in Sec.~\ref{sec:modes}, this symmetry may be broken as, e.g., the excitation can be localized on the top or bottom layer, or there can be different level shifts on each layer. 
 
\section{Bilayer array}
\subsection{Collective modes}
\label{sec:modes}

Although in our numerical analysis we solve the full system of Eqs.~\eqref{eq:eom},
the response of the lattice can be understood by considering the collective eigenmodes $\vec{v}_n$ of $\mathcal{H}$, and their eigenvalues $\delta_n+i\upsilon_n$, where $\delta_n$ is the collective level shift and $\upsilon_n$ is the collective linewidth~\cite{Rusek96,JenkinsLongPRB,Jenkins2012a,Castin13,
Jenkins_long16,Lee16}. For the applications that we consider in this paper,
the bilayer lattice has four collective eigenmodes of interest. For the $y$-polarized incident field, there is a symmetric, $y$-polarized mode $\vec{v}_{ys}$ with amplitude $\Pc_{ys}$, where $\Pc_n = \vec{v}_n^T b$ (since $\mathcal{H}$ is symmetric $\vec{v}_j^T\vec{v}_k=\delta_{jk}$\footnote{For some modes, $\vec{v}_j^T \vec{v}_j=0$ is also possible, but these are not encountered in the current system.}).
This mode consists of all dipoles oscillating uniformly and in-phase in the $y$ direction (in the plane of the lattice). There is then also an antisymmetric mode, with amplitude $\Pc_{ya}$, where all dipoles within a layer oscillate in phase, but a $\pi$ phase difference exists between the top and bottom layers, i.e.\ $\Pc_y(x=a/2)=-\Pc_y(x=-a/2)$. These modes are coupled to the out-of-plane modes,
one symmetric with amplitude $\Pc_{xs}$, with uniform distribution of dipoles in the $x$ direction, and an antisymmetric one with amplitude $\Pc_{xa}$, with a $\pi$ phase difference between the two layers. For a subwavelength lattice there is only the zero-order Bragg peak for scattering which becomes infinitely sharp in the limit of infinite lattice size~\cite{CAIT,Facchinetti16,Facchinetti16,Javanainen19,Ballantine20ant}. Because no emission occurs along the dipole axis, the out-of-plane modes cannot scatter in this direction, and so are strongly subradiant, with $\lim_{N\rightarrow \infty} \upsilon_{xs}, \upsilon_{xa} = 0$ for infinite atom number $N$.

\begin{figure}[t]
 \centering
  \includegraphics[width=\columnwidth]{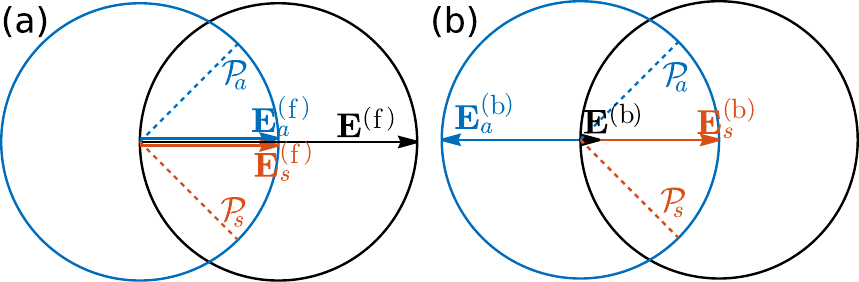}
 \caption{Interference of light in forward and backward directions due to differing mode symmetry, for $a=\lambda/4$. (a) Mode amplitude and scattered field amplitudes in complex plane. If $\Pc_s$ is $\pi/2$ out of phase with $\Pc_a$, then the scattered fields (shown here up to an irrelevant global phase factor) interfere constructively, leading to the total forward scattered field $\vec{E}^{(f)}=\vec{E}_s^{(f)}+\vec{E}_a^{(f)}$. If the phase of $\Pc_a$ varies, while keeping $\Pc_s$ fixed, then $\vec{E}_a^{(f)}$ will follow the blue circle, and $\vec{E}^{(f)}$ the black circle. (b) Conversely, $\vec{E}_a^{(b)}=-\vec{E}_a^{(f)}$, and so destructive interference occurs for the relative phase $\Pc_a=i\Pc_s$, with $\vec{E}^{(b)}=\vec{E}_s^{(b)}+\vec{E}_a^{(b)}=0$. As the relative phase varies and the forward scattering decreases the backward scattering correspondingly increases. }
 \label{fig:sym}
\end{figure}

It is the differing symmetry of these modes which allows full control of the directionality of both absorption and emission. As described in Sec.~\ref{sec:mono}, this is prevented in the single-layer case by the symmetry between the forward and backward directions. 
Writing the total field $\vec{E}_{\rm tot}^{(f,b)}=\cbE^{(f,b)}+\vec{E}^{(f,b)}$ propagating in the forward ($f$) or backward ($b$) direction as the sum of the incident and scattered field in the same directions, the scattered field can further be decomposed into $\vec{E}^{(f,b)}=\vec{E}_s^{(f,b)}+\vec{E}_a^{(f,b)}$, with $\vec{E}_s^{(f,b)}$ ($\vec{E}_a^{(f,b)}$) the contribution from the symmetric (antisymmetric) mode. For the bilayer lattice, the symmetric mode scatters equally in both directions, with even parity,
\begin{equation}
\vec{E}_s^{(f)} = \vec{E}_s^{(b)},
\end{equation}
but the antisymmetric mode exhibits odd parity and the scattered field in the forward direction is the negative of that in the backward direction,
\begin{equation}
\vec{E}_a^{(f)} = -\vec{E}_a^{(b)}.
\end{equation}
Hence, when both modes are excited, the interference between the different contributions in both directions will lead to different total field amplitudes, breaking the symmetry and allowing for both emission and absorption even of unidirectional-propagating light. 
The antisymmetric mode does not generate the conventional half-wave loss of the field near the atom array interface that could find useful applications in producing strong near fields.

For incident light in the forward direction, with $\alpha=1$ in Eq.~(\ref{eq:direction}), the symmetric mode will be driven with an amplitude $\cos{(ka/2)}$ and the antisymmetric mode with an amplitude $i\sin(ka/2)$. 
While high-efficiency absorption can be achieved for arbitrary $a$, this is most simply done for $a=(2n+1)\lambda/4$, where both symmetric and antisymmetric modes can be equally excited without extraneous level shifts. We first describe the concept of suppressed scattering during absorption with $a=\lambda/4$.

The relative phase of scattering is illustrated in Fig.~\ref{fig:sym}. 
The arrows show the case of unidirectional absorption or emission, where $\Pc_a=i\Pc_s$, such that $\vec{E}_s^{(f)}=\vec{E}_a^{(f)}$ and $\vec{E}_s^{(b)}=-\vec{E}_a^{(b)}$. A different phase of $\Pc_a$ relative to $\Pc_s$ will lead to a different phase of $\vec{E}_a^{(f,b)}$ relative to $\vec{E}_s^{(f,b)}$ and the interference will, in general, be only partial, with $\Pc_a=-i\Pc_s$ corresponding to the opposite case of full destructive interference in the forward direction and constructive in the backward. This description in terms of symmetric and antisymmetric modes is also equivalent to a description of interference between scattering from the top and bottom layers. 

The bilayer can also be considered as a regular lattice of two-atom unit cells, each consisting of a pair of atoms at $\pm a/2\unitvec{x}$. 
The multipole composition of each such unit cell can be found by expanding the scattered field,
\begin{equation}
\vec{E} = \sum_{l=0}^\infty\sum_{m=-l}^{l}\left( \alpha_{E,lm}\vecg{\Psi}_{lm}+\alpha_{B,lm}\vecg{\Phi}_{lm}\right),
\end{equation}
where $\vecg{\Psi}_{lm}$, $\vecg{\Phi}_{lm}$ are the vector spherical harmonics and $\alpha_{E,lm}$ ($\alpha_{B,lm}$) are electric (magnetic) dipole, quadrupole, etc., coefficients for $l=1,2,\ldots$, respectively~\cite{Jackson,Grahn12}.
Then each of the four collective eigenmodes can be seen as a simple repetition of one of the modes of the single unit cell. $\Pc_{ys}$ corresponds to a uniform repetition of two equal in-phase dipoles that generate a strong electric dipole, and $\Pc_{ya}$ to a uniform repetition of two dipoles of equal magnitude but a $\pi$ phase difference that produces magnetic dipole and electric quadrupole contributions.
More complex non-Bravais lattices, for example with square or diamond unit cells, can be used to further refine and control these contributions~\cite{Ballantine20Huygens,Alaee20}, as discussed in Sec.~\ref{sec:hs}. In general, the multipole scattering can be separated into symmetric (electric dipole, magnetic quadrupole, etc.) and antisymmetric (magnetic dipole, electric quadrupole, etc.), when only the overall symmetry is required to engineer the desired directional constructive or destructive interference~\cite{Liu17mm}.

While the in-plane modes couple strongly to the incident $y$-polarized field, the out-of-plane subradiant modes consist of atomic dipoles oscillating perpendicular to the light polarization direction.
We will generalize the techniques of Refs.~\cite{Facchinetti16,Facchinetti18} to a multimode case to couple different spatially delocalized collective modes to each other using an induced level shift for the atoms. 
The overall response of the lattice can then be understood in terms of a simple four-mode-model (see Appendix~\ref{sec:fmm}). A differential level shift between the two layers, i.e.\ a different shift of the $J=0\rightarrow J^\prime =1$ spacing on each layer while keeping the $J^\prime=1, m=\pm 1$ levels degenerate, is 
encapsulated by the effective parameter $\Delta_{a,s}$ in Eq.~(\ref{eq:bd2}). This differential level shift breaks the symmetry between the layers and couples each symmetric mode to the corresponding antisymmetric mode. An overall linear Zeeman splitting of the $m=\pm 1$ levels with identical values on each layer, meanwhile, is encapsulated by the parameter $\Delta_{x,y}$ in Eq.~(\ref{eq:bd1}). This Zeeman splitting breaks the isotropy of the transition, causing the dipoles to rotate and thereby coupling the in-plane and out-of-plane modes. The required differential level shifts and Zeeman splittings could both be achieved by ac Stark shifts of suitably polarized lasers or microwaves~\cite{gerbier_pra_2006}, allowing for much faster switching than possible with magnetic fields.

\subsection{Photon absorption}
\label{sec:abs}

\begin{figure}[t]
  \centering
   \includegraphics[width=\columnwidth]{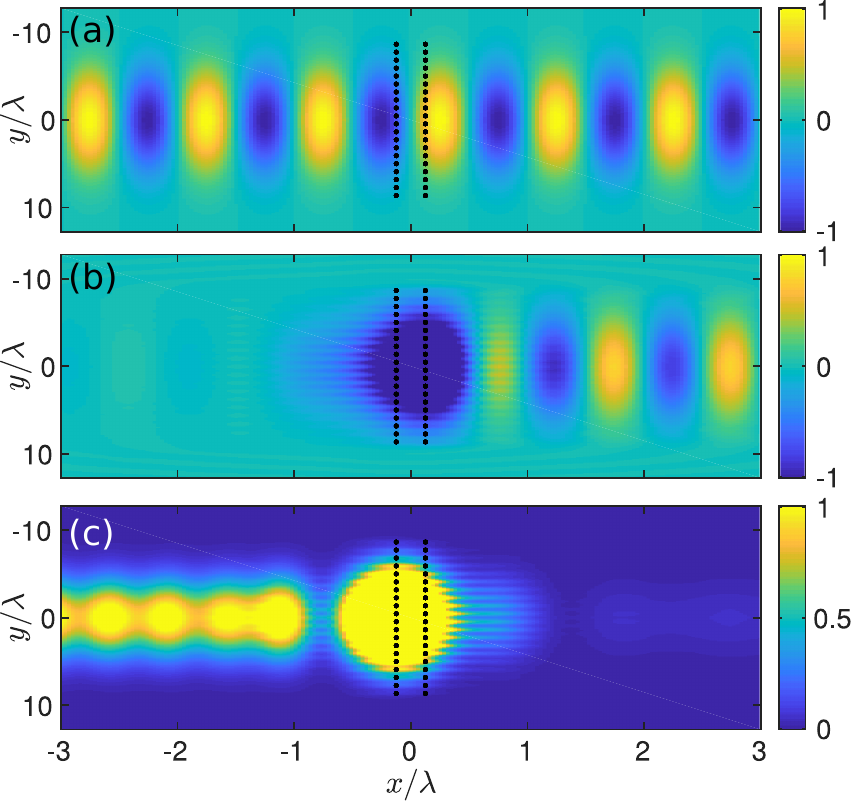}
  \caption{Suppression of transmission while photon absorption is underway. (a) Spatial distribution of real part ${\rm Re}({\cal E})$, propagating from left to right. (b) Real part of scattered field ${\rm Re}(E_y)$. Destructive interference occurs in backward direction leading to scattering only in forward direction, which is $\pi$ out-of-phase with incident light.  (c) Total intensity $|\vec{E}_{\rm tot}|^2$ at $t=0$. Incident light from the left is absorbed by the array, while for $x>0$ incident field seen in (a) is canceled by scattered field in (b). $20\times 20\times 2$ lattice illustrated by black dots, $d=0.91\lambda$, $a=0.25\lambda$.  }
  \label{fig:gspace}
\end{figure}

\begin{figure}[t]
 \centering
  \includegraphics[width=\columnwidth]{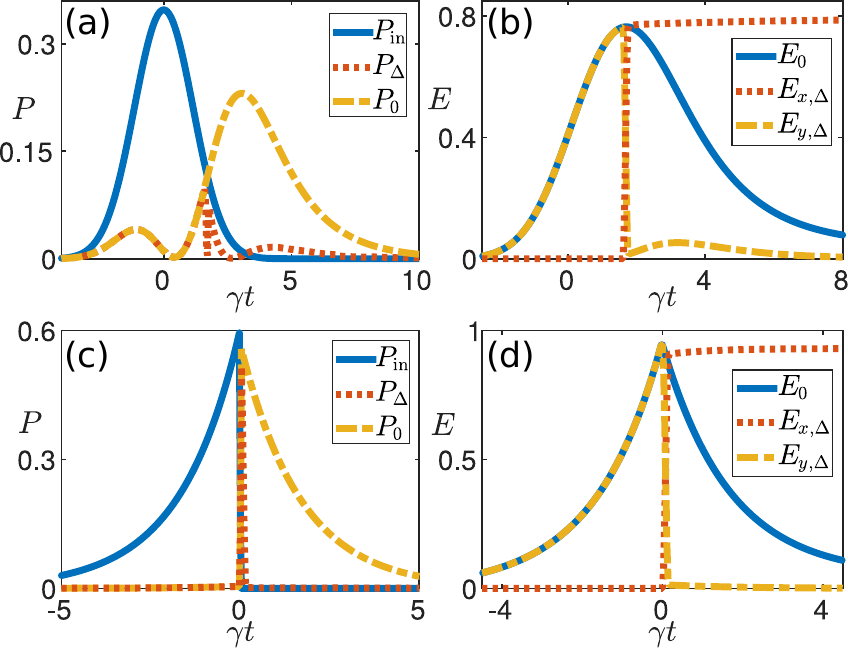}
 \caption{Absorption of single photon incident on bilayer from one direction. (a) Total incident power ($P_{\rm in}$) and total transmitted power pulse which is Gaussian in space ($w_0=5.8\lambda$) and time ($\gamma\tau=2.3$), when Zeeman splitting is not present ($P_{0}$), and when coupling $\Delta_{x,y}=10\gamma$ is turned on briefly at $\gamma t=1.6$ ($P_{\Delta}$), in units of $\gamma\hbar\Omega$. (b) Corresponding energy in units of $\hbar\Omega$ of all modes in absence of Zeeman splitting ($E_0$), in-plane modes with Zeeman splitting ($E_{y,\Delta}$), and out-of-plane modes with Zeeman splitting ($E_{x,\Delta}$). The coupling is turned off after a period $0.15/\gamma$ to store excitation. (c) Power and (d) stored energy for pulse which is exponential in time, $\tau=0.3\gamma$, with and without $\Delta_{x,y}=10\gamma$ from $t=0$ to $t=0.15\gamma$. $20\times 20\times 2$ lattice with $d=0.91\lambda$, $a=0.25\lambda$.  }
 \label{fig:abs}
\end{figure}

We consider absorption of a time-dependent forward-propagating single-photon pulse, with Gaussian time profile $h(t)=\exp{(-t^2/\tau^2)}$ in Eq.~(\ref{eq:drive}) and $\alpha=1$ in Eq.~(\ref{eq:direction}).
The incident $y$ polarized field will drive a combination of the in-plane modes $\Pc_{ys}$ and $\Pc_{ya}$. We first treat the case $a=\lambda/4$, where both in-plane modes experience equal driving from the incident field (see Appendix~\ref{sec:fmm}), leading to equal excitation. 
For this value of $a$, we find the collective line shifts, shown in Fig.~\ref{fig:fig1}, intersect at $d\simeq 0.91\lambda$. Although we will show later how the photon absorption process is not very sensitive to the values of $d$, this condition maximizes efficiency by ensuring that each mode oscillates at the same resonance frequency, therefore keeping the relative phase constant during the absorption process. For $d\simeq 0.91\lambda$, the antisymmetric in-plane mode $\Pc_{ya}$ has a normalized magnetic dipole contribution $|\alpha_{B,10}|^2\simeq 0.63$ from each unit cell and an electric quadrupole contribution $\sum_m|\alpha_{E,2m}|^2\simeq 0.37$.

Exciting the atomic array leads to scattering as the process is ongoing. The spatial distribution of the intensities of the incident, scattered, and total fields are shown in Fig.~\ref{fig:gspace}, for peak incident field amplitude. 
Scattering from $\Pc_{ya}$ and $\Pc_{ys}$, shown in Fig.~\ref{fig:gspace}(b) interfere destructively in the backward direction leading to zero reflection. While they interfere constructively in the forward direction, they are out of phase with the incident light, meaning transmission is also suppressed, as seen in Fig.~\ref{fig:gspace}(c). The time dynamics of the process is shown in Fig.~\ref{fig:abs}. In the absence of shifts between the $m=\pm1$ levels, the scattering would continue as the incident pulse fell away, and the pulse would simply radiate forward at a delayed time, as shown in Fig.~\ref{fig:abs}(a). We will suppress this scattering by coupling collective modes to each other~\cite{Facchinetti16}.
At the point of maximum mode occupation a linear Zeeman splitting $\Delta_{x,y}$ that breaks the degeneracy between the  $m=\pm1$ levels can be applied uniformly to all atoms. This causes the polarization to rotate out of the plane, coupling $\Pc_{ya}$ to $\Pc_{xa}$, and $\Pc_{ys}$ to $\Pc_{xs}$, respectively (see Appendix~\ref{sec:fmm}), as the uniform coupling preserves the symmetry between layers. The level shifts are then turned off, and the excitation is stored in the subradiant modes, with $\upsilon_{xa}\approx 0.001\gamma$ and $\upsilon_{xs}\approx 0.03\gamma$ for the parameters considered here.
The result is shown in Fig.~\ref{fig:abs}(a,b), with photon absorption efficiency 0.79.   

Although the storage mechanism is valid for general lattice separations, the process is simplest to describe for the case of large spacing between the two layers in terms of 1D electrodynamics~\cite{Ruostekoski17}, as is done in Appendix~\ref{sec:fmm}.
When the separation $\lambda\lesssim a \ll \sqrt{A}$, where $A$ is the area of each layer, propagation between layers is effectively 1D~\cite{dalibardexp,Javanainen17,Facchinetti18,Javanainen19}. Then, each layer is equivalent to a single ``superatom'' with modified collective linewidth $\upsilon_{\rm 1D}=3\pi\gamma/(k^2 d^2)$
of the uniformly excited array~\cite{CAIT}, coupled by a 1D channel~\cite{Javanainen1999a,Ruostekoski17}. Subradiant states have similarly been shown to be accessible in a 1D system of waveguide-coupled atoms by varying the detuning of each atom~\cite{Ruostekoski17}.

The storage efficiency can be improved by understanding the time-reversal invariance. If the in-plane modes were allowed to decay, they would do so exponentially, and the resulting intensity would also have an exponential profile, differing significantly from the Gaussian input. In order to better match the time-reversed process, we consider also an exponential pulse, $h(t)=\exp{(t/\tau)}\Theta(-t)$, where $\Theta(t)$ is the Heaviside function. The result is shown in Fig.~\ref{fig:abs}(c,d), with a higher efficiency of 0.93. 

While $d\simeq 0.91\lambda$ results in optimal absorption for $a=0.25\lambda$ with equal collective mode resonances, small variations in $d$ lead only to a small reduction in the efficiency. For example, we find for $d=0.8\lambda$ an efficiency $\approx 0.7$ for Gaussian input, rising to $\approx 0.79$ for an exponential input pulse. The efficiency could be further improved by varying the level shifts in time during the absorption~\cite{Rubies21}. However, the differing collective line shifts, $\delta_{xs}\neq\delta_{xa}$,  mean that each mode evolves at a different frequency, complicating the effect of varying the level shifts over longer timescales.

High-efficiency absorption is also possible for arbitrary $a\neq (2n+1)\lambda/4$. To achieve this, we introduce a difference in the level shifts between the layers at $x=\pm a/2$, as could be achieved by the ac Stark shift of light with an intensity gradient in the $x$ direction. 
Upon introducing different level shifts between each layer, the modes $\Pc_{ys}$ and $\Pc_{ya}$ of $\mathcal{H}$ are no longer eigenmodes of $\mathcal{H}+\delta{H}$, but are coupled together with nonzero $\Delta_{a,s}$ in Eq.~(\ref{eq:bd2}) (see Appendix~\ref{sec:fmm}). 
For general $a$, $\upsilon_a\neq \upsilon_s$, and to suppress backwards scattering, with $\vec{E}_s^{(b)}=-\vec{E}_a^{(b)}$, we excite a linear combination with unequal amplitudes $|\Pc_s|\neq|\Pc_a|$. The desired combination can be targeted by choosing $\Delta_{a,s}$ appropriately. Again, turning on a linear Zeeman splitting $\Delta_{x,y}$, that breaks the degeneracy between the  $m=\pm1$ levels at the point of maximum occupation, transfers the excitation to a corresponding linear combination of $\Pc_{xs}$ and $\Pc_{xa}$, where light can be stored. 
We briefly consider a numerical example with $a=0.9\lambda$. For this separation there is no choice of $d$ such that $\delta_{ys}=\delta_{ya}$. Despite this, we still find a high efficiency $\approx 0.7$ can be achieved for a Gaussian input pulse, rising to $\approx 0.85$ when the incident pulse is tailored to match the time-reversal of emitted light, both for $d=0.9\lambda$.

\subsection{Photon retrieval}

\begin{figure}[t]
 \centering
  \includegraphics[width=\columnwidth]{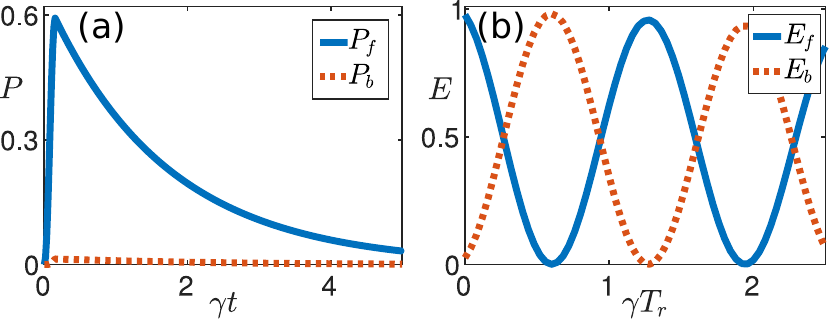}
 \caption{Retrieval of the photon from storage. (a) Power $P_f$ ($P_b$) in the forward (backward) direction, in units $\hbar\gamma\Omega$, when the photon is released by turning on level shifts $\Delta_{x,y}=10\gamma$ from $T_r=0$ until $\gamma t=0.15$. The excitation is transferred to a superposition of the in-plane modes where it decays exponentially. $97\%$ of the total energy is radiated in the forward direction. (b) Total integrated energy $E_f$ ($E_b$) in units of $\hbar\Omega$ radiated in the forward (backward) direction, when the excitation is released at time $T_r$. Lattice parameters as in Fig.~\ref{fig:abs}.
  }
 \label{fig:rel}
\end{figure}

The photon can be released after some time by re-activating the level shifts, coupling $\Pc_{xa}$ to $\Pc_{ya}$, and $\Pc_{xs}$ to $\Pc_{ys}$, respectively, and allowing the excitation to radiate.
As noted in Sec.~\ref{sec:abs}, if the photon is stored in a superposition of $\Pc_{xs}$ and $\Pc_{xa}$, then each mode will evolve with a different frequency, $\delta_{xs}$ and $\delta_{xa}$, respectively. 
Hence, the phase difference between the two modes will vary depending on the time, $T_r$, at which the photon is released. This can be used to control the direction of emission, allowing the photon to be released in the forward direction, backward direction, or an arbitrary fraction in each, as the interference depends on the relative phase. The resulting total integrated power transmitted in each direction, when the level shifts $\Delta_{x,y}$ are turned on at a time $T_r$, is shown in Fig.~\ref{fig:rel}, for the case $a=0.25\lambda$ and $\Delta_{a,s}=0$.

\subsection{Storage and coherent oscillations}
\label{sec:storage}

\begin{figure}[t]
 \centering
  \includegraphics[width=\columnwidth]{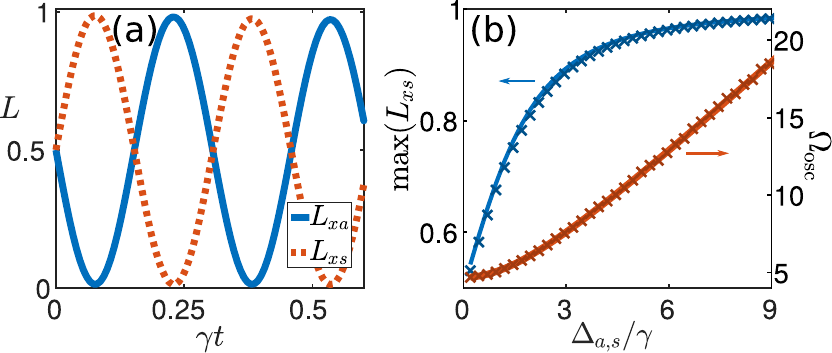}
 \caption{Coherent oscillations between two subradiant modes. (a) Occupation of the symmetric and antisymmetric out-of-plane modes when they are coupled by $\Delta_{a,s}=10\gamma$. (b) Maximum occupation of the symmetric mode $L_{xs}$ (left axis), and frequency of oscillation $\Omega_{\rm osc}$ (right axis), for different coupling strengths $\Delta_{a,s}$. Solid lines in (b) are derived from four-mode model. Data in (a) and markers in (b) correspond to numerical simulation for lattice parameters as in Fig.~\ref{fig:abs}.
  }
 \label{fig:Losc}
\end{figure}

As the emission is the time-reversal of absorption, the converse is also true; an incident photon in an arbitrary combination of left and right incidence can be absorbed, leading to a specific amplitude and phase relationship of $\Pc_{xa}$ and $\Pc_{xs}$. This storage can be further controlled by coupling the two out-of-plane modes. A differential level shift between the layers again gives nonzero $\Delta_{a,s}$ in Eq.~(\ref{eq:fourmm}). This couples $\Pc_{xs}$ to $\Pc_{xa}$, leading to coherent oscillations between the two populations. The resulting mode occupations can be defined as~\cite{Facchinetti16}
\begin{equation}
L_j = \frac{|\vec{v}_j^T\vec{b}|^2}{\sum_k |\vec{v}_k^T\vec{b}|^2},
\end{equation}
where $j$ indexes all collective modes $\vec{v}_j$ of ${\cal H}$, as described in Sec.~\ref{sec:modes}. The occupations of the two out-of-plane modes are shown in Fig.~\ref{fig:Losc}(a), with $L_j\approx 0$ for all other modes. Coupling $\Pc_{x,a}$ and $\Pc_{x,s}$ thus allows for the occupation to be redistributed in either a single mode or an equal combination. An analytical solution of Eq.~(\ref{eq:fourmm}) shows that the oscillations have a frequency $\Omega_{\rm osc}=2\sqrt{\Delta_{a,s}^2+(\delta_{xs}-\delta_{xa})^2}$, while the maximum occupation of, e.g., $\Pc_{xs}$ is $1/2+\Delta_{a,s}/\Omega_{\rm osc}$, when starting from a symmetric superposition of both modes, as illustrated by the solid lines in Fig.~\ref{fig:Losc}(b).

\section{Square unit cells}
\label{sec:hs}

\begin{figure}[t]
 \centering
  \includegraphics[width=\columnwidth]{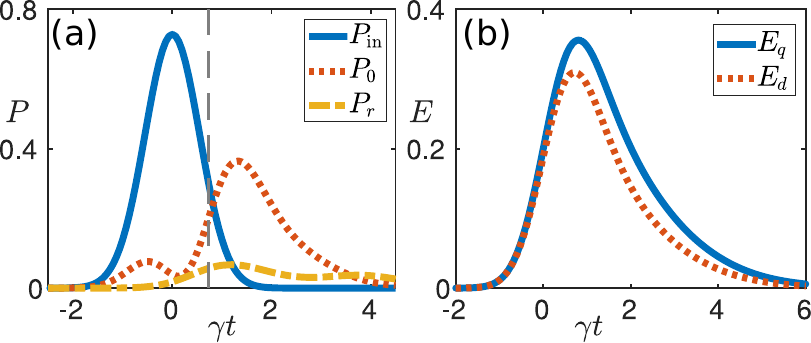}
 \caption{Absorption of a Gaussian pulse, with $\gamma\tau=2.3$, by a square-unit-cell bilayer. (a) Incident power ($P_{\rm i}$), transmitted power ($P_0$) and reflected power ($P_r$) in units of $\gamma\hbar\Omega$. (b) Energy stored in the antisymmetric electric quadrupole mode ($E_q$), and the symmetric electric dipole mode ($E_d$), in units of $\hbar\Omega$. Here the absorption and emission proceed without level shifts. The point of maximum occupation is marked by the dashed line in (a). $19\times 19\times 4$ lattice with $d=0.83\lambda$, $a=0.25\lambda$. }
 \label{fig:sq2}
\end{figure}

In order to show how the symmetry principles of the scattering are more general, we consider the more complex square-unit-cell bilayer lattice. Here each unit cell contains four atoms in a square of side-length $a$, oriented in the $xy$ plane, with the 2D lattice of unit cells with lattice constant $d$ extended in the $yz$ plane. The collective modes of the lattice can be understood as combinations of the modes of individual unit cells, which can themselves be decomposed into various electric dipole, magnetic dipole, electric quadrupole, etc., contributions~\cite{Ballantine21quantum,Ballantine20Huygens}.

Again, the modes fall into a symmetric family (electric dipole, magnetic quadrupole, etc.), and an antisymmetric family (magnetic dipole, electric quadrupole, etc.). Exciting one mode from each family can suppress transmission while the pulse is absorbed, but to achieve a high-fidelity storage and retrieval requires overlapping mode resonances. We find a symmetric and antisymmetric collective mode with similar line shifts for $d\simeq 0.83\lambda$, $a\simeq 0.25\lambda$, for which the electric dipole and quadrupole modes have $\delta_n+i\upsilon_n\simeq 1.6+0.66 i$ and $1.6+0.59 i$, respectively. These modes exhibit the same symmetry in scattering as the symmetric and antisymmetric modes, $\Pc_{ys}$ and $\Pc_{ya}$, of the simple bilayer lattice, with $\vec{E}_d^{(f)}=\vec{E}_d^{(b)}$ and $\vec{E}_q^{(f)}=-\vec{E}_q^{(b)}$, where $\vec{E}_d^{(f,b)}$ and $\vec{E}_q^{(f,b)}$ are the electric dipole and electric quadrupole scattered fields, respectively, in the $\pm \unitvec{x}$ directions. As in Sec.~\ref{sec:modes}, this symmetry allows destructive interference in the reflected light, and constructive interference in the forward scattered light suppressing transmission.  The principle is the same as demonstrated in Figs.~\ref{fig:sym} and~\ref{fig:gspace}.

The resulting absorption of a uni-directional Gaussian pulse $h(t)=\exp(-t^2/\tau^2)$ with $\tau=1.1\gamma$ is shown in Fig.~\ref{fig:sq2}. In this case, the pulse is absorbed and re-emitted in the absence of any externally applied level shifts. The storage of the pulse is not trivial, as polarization consists of a combination of electric-dipole and electric-quadrupole modes. The polarization will, in general, be a mix of $x$ and $y$ polarization with a different direction on each atom (see Ref.~\cite{Ballantine20Huygens} Supplemental Material). However, at the time of maximum mode occupation, $\approx 0.75$ of the incident energy has been transferred to the atomic excitations.

\section{Concluding remarks}

Our protocols provide full control over the directional absorption, storage, and emission of a single-photon pulse. Light incident from the forward direction, backward direction, or an arbitrary combination can first be absorbed and stored by coupling the in-plane to the out-of-plane modes. The excitation can then be transferred to any desired combination of spatially delocalized out-of-plane collective modes by coupling symmetric and antisymmetric modes, e.g., to store it in the more subradiant antisymmetric mode $\Pc_{xa}$. After some time, it can then be transferred to a superposition of the antisymmetric and symmetric out-of-plane modes $\Pc_{xa}$ and $\Pc_{xs}$, and, by choosing a suitable release time $T_r$, can be emitted in an arbitrarily weighted superposition of forward and backward directions, independent of that of the initial incident photon pulse. Such efficient and controllable absorption, storage, and emission could be utilized to take advantage of the effective 1D emission of atomic lattices, as part of a larger architecture, to transfer excitations forward and backward between layers for processing or storage. Additional control of the direction, phase, and polarization of the emitted photon could also be achieved with more complex, spatially varying level shifts~\cite{Ballantine21wavefront}, with the system obtaining more general properties of highly sought after quantum metasurfaces~\cite{Solntsev21}.

 \begin{acknowledgments}
We acknowledge financial support from the United Kingdom Engineering and Physical Sciences Research Council (EPSRC) (Grants No. EP/S002952/1, No. EP/P026133/1, and No. EP/W005638/1).
\end{acknowledgments}

\appendix
\section{Four-mode model}
\label{sec:fmm}

In the main text we make use of a four-mode model to explain the dynamics in terms of symmetric and antisymmetric in-plane and out-of-plane modes, with various couplings between them. Here, we illustrate how these couplings arise.

We first consider coupling the in-plane component to the out-of-plane component. For a single atom at $\vec{r}=0$ in the circular polarization basis the polarization amplitudes evolve as
\begin{equation}
\partial_t\Pc_{\mu} = (i\Delta_{\mu}+i\delta\Omega-\gamma)\Pc_{\mu} + \zeta_\mu(t), 
\end{equation}
with $\zeta_\mu(t)=i(\xi/\mathcal{D})\unitvec{e}_\mu^{\ast}\cdot \epsilon_0\cbE(0,t)$.
Writing these amplitudes in the Cartesian basis gives 
\begin{align*}
\partial_t \Pc_x &= (i\tilde{\delta}+i\delta\Omega-\gamma)\Pc_x +\bar{\delta}\Pc_y, \\
\partial_t \Pc_y &= (i\tilde{\delta}+i\delta\Omega-\gamma)\Pc_y -\bar{\delta}\Pc_x + \zeta_y(t), 
\end{align*}
with $\tilde{\delta}=(\Delta_{+}+\Delta_{-})/2$, and $\bar{\delta}=(\Delta_{+}-\Delta_{-})/2$ coupling the $x$ and $y$ polarization on a single atom, and where the $y$ polarized incident field propagating in the $x$ direction [see Eq.~(\ref{eq:drive})] drives only $\Pc_y$.
The same principle of coupling can be generalized to a single 2D array of atoms in the $yz$ plane in which case the dipoles oscillating in phase in the $y$ direction correspond to the in-plane mode and
the dipoles oscillating in phase in the $x$ direction correspond to the out-of-plane mode~\cite{Facchinetti16,Facchinetti18}.
Similarly for the full bilayer lattice the in-plane and out-of-plane modes of the unperturbed lattice are no longer eigenmodes of the system in the presence of Zeeman splitting but are coupled together. Coupling between, e.g., the symmetric modes can be described by a two-mode model in analogy with the single-atom case,
\begin{equation*}
\setlength\arraycolsep{-5pt}
\setlength\arraycolsep{-5pt}
\begin{pmatrix}
\dot{\Pc}_{xs} \\ \dot{\Pc}_{ys}
\end{pmatrix} = i\begin{pmatrix}
\lambda_{xs}+\delta\Omega+\tilde{\delta} & -i\bar{\delta} \\ i\bar{\delta} & \lambda_{ys}+\delta\Omega+\tilde{\delta}
\end{pmatrix}\begin{pmatrix}
\Pc_{xs} \\ \Pc_{ys}
\end{pmatrix} + \begin{pmatrix}
0 \\ \zeta_{ys} 
\end{pmatrix},
\end{equation*}
where $\lambda_n=\delta_n+i\upsilon_n$ and
\begin{equation}
\zeta_{ys} = i\sqrt{2}(\xi/\mathcal{D})\epsilon_0 \mathcal{N} h(t)  \cos{(ka/2)}.
\end{equation}
An identical equation describes coupling between $\Pc_{ya}$ and $\Pc_{xa}$, with $\lambda_{ys}$, $\lambda_{xs}$ replaced with $\lambda_{ya}$, $\lambda_{xa}$ and $\zeta_{ys}$ replaced with
\begin{equation}
\zeta_{ya} = -\sqrt{2}(\xi/\mathcal{D})\epsilon_0 \mathcal{N} h(t) \sin{(ka/2)}.
\end{equation}

Coupling between, e.g., the symmetric and antisymmetric in-plane modes is most easily explained by first considering the regime $\lambda\lesssim a \ll \sqrt{A}$, where $A$ is the area of each layer. 
In this regime analytic expressions can be found for linewidths and the relative line shifts of the modes. However, the case $a<\lambda$ is described by identical couplings between modes, but with $\lambda_{ya}$, $\lambda_{ys}$ replaced with values found from full numerical analysis. 

For $\lambda\lesssim a\ll \sqrt{A}$, light propagation between the layers is effectively 1D with each lattice responding as a single ``superatom''~\cite{Facchinetti18,Javanainen19}. We write the modes in terms of the uniform in-plane collective excitations $\Pc_{y1}$ of layer 1 (at $x=-a/2$) and $\Pc_{y2}$ of layer 2 (at $x=a/2$), respectively, as
\begin{align}
\Pc_{ys} = \frac{1}{\sqrt{2}}\left(\Pc_{y1}+\Pc_{y2}\right), \\
\Pc_{ya} = \frac{1}{\sqrt{2}}\left(\Pc_{y1}-\Pc_{y2}\right).
\end{align}
We apply uniform level shifts
$\Delta^{(j)}_\mu=\Delta^{(1)}$ for $j\leq N/2$, with $N$ total atoms ordered such that $j\leq N/2$ refers to the bottom ($x<0$) layer, and $\Delta^{(j)}_\mu=\Delta^{(N)}$ for $j> N/2$, referring to the top ($x>0$) layer. Then the amplitudes of each layer evolve as 
\begin{multline*}
\setlength\arraycolsep{-5pt}
\begin{pmatrix}
\dot{\Pc}_{y1} \\ \dot{\Pc}_{y2}
\end{pmatrix} = i\begin{pmatrix}
\lambda_{\rm 1D}+\delta\Omega+ \Delta^{(1)} & i\mathsf{G}_{12} \\ i\mathsf{G}_{12} & \lambda_{\rm 1D}+\delta\Omega +  \Delta^{(N)}
\end{pmatrix} 
\begin{pmatrix}
\Pc_{y1} \\ \Pc_{y2}
\end{pmatrix} \\
+\begin{pmatrix}
\zeta_{y1} \\ \zeta_{y2}
\end{pmatrix},
\end{multline*}
where $\lambda_{\rm 1D}=\delta_{\rm 1D}+i\upsilon_{\rm 1D}$ consists of a collective level shift $\delta_{\rm 1D}$ and linewidth $\upsilon_{\rm 1D}=3\pi\gamma/(k^2 d^2)$ of the uniformly excited array~\cite{CAIT}, and $\mathsf{G}_{12}=\upsilon_{\rm 1D}\exp{(ika)}$ is an effective 1D interaction between the layers~\cite{Ruostekoski17}. Returning to the basis  of $\Pc_{ya}$, $\Pc_{ys}$, these are no longer eigenmodes but are coupled together, 
\begin{equation*}
\partial_t\begin{pmatrix}
\Pc_{ya} \\ \Pc_{ys}
\end{pmatrix} = i\begin{pmatrix}
\lambda_{ya}+\delta\Omega+\tilde{\delta} && \Delta_{a,s} \\ \Delta_{a,s} && \lambda_{ys}+\delta\Omega+\tilde{\delta} 
\end{pmatrix} +\begin{pmatrix}
\zeta_{ya} \\ \zeta_{ys}
\end{pmatrix},
\end{equation*}
with $\lambda_{ya} = \lambda_{\rm 1D}-i\mathsf{G}_{12}$, $\lambda_{ys} = \lambda_{\rm 1D}+i\mathsf{G}_{12}$, $\tilde{\delta}=(\Delta^{(1)}+\Delta^{(N)})/2$, and $\Delta_{a,s}=(\Delta^{(1)}-\Delta^{(N)})/2$.

Now we take the level shifts to be equal in each plane of the bilayer array  (no variation in the $yz$ plane), with both a possible shift between the planes (along the $x$ direction) and a possible Zeeman splitting $\Delta_{+}^{(j)}\neq\Delta_{-}^{(j)}$ between the $m=\pm1$ levels of each atom. Combining the transformation to Cartesian coordinates with that to a diagonal basis of $\mathcal{H}$, the full control can then be described by a four-mode model;
\begin{equation}
\label{eq:fourmm}
\partial_t\begin{pmatrix}
\Pc_{xa} \\ \Pc_{ya} \\ \Pc_{xs} \\ \Pc_{ys}
\end{pmatrix}=i(\Lambda +\delta\Omega\idop +\delta\mathcal{H})\begin{pmatrix}
\Pc_{xa} \\ \Pc_{ya} \\ \Pc_{xs} \\ \Pc_{ys}
\end{pmatrix}+\begin{pmatrix}
0 \\ \zeta_{ya} \\ 0 \\ \zeta_{ys}
\end{pmatrix},
\end{equation}
with 
$\Lambda = \mathrm{diag}(\lambda_{xa},\lambda_{ya},\lambda_{xs},\lambda_{ys})$ for $\lambda_n=\delta_n+i\upsilon_n$ and 
\begin{equation}
\label{eq:dh}
\setlength\arraycolsep{1pt}
\delta\mathcal{H}=\begin{pmatrix}
\tilde{\delta} & -i\Delta_{x,y} & \Delta_{a,s} & -i\Delta_{a,s;x,y} \\
i\Delta_{x,y} & \tilde{\delta} & i\Delta_{a,s;x,y} & \Delta_{a,s} \\
\Delta_{a,s} & -i\Delta_{a,s;x,y} & \tilde{\delta} & -i\Delta_{x,y} \\
i\Delta_{a,s;x,y} & \Delta_{a,s} & i\Delta_{x,y} & \tilde{\delta}
\end{pmatrix},
\end{equation}
where
\begin{align}
\label{eq:td1}
\tilde{\delta} &= (\Delta_{+}^{(1)}+\Delta_{-}^{(1)}+\Delta_{+}^{(N)}+\Delta_{-}^{(N)})/4, \\
\label{eq:bd1}
\Delta_{x,y} &= (\Delta_{+}^{(1)}-\Delta_{-}^{(1)}+\Delta_{+}^{(N)}-\Delta_{-}^{(N)})/4, \\
\label{eq:bd2}
\Delta_{a,s} &= (\Delta_{+}^{(1)}+\Delta_{-}^{(1)}-\Delta_{+}^{(N)}-\Delta_{-}^{(N)})/4, \\
\label{eq:td2}
\Delta_{a,s;x,y} &= (\Delta_{+}^{(1)}-\Delta_{-}^{(1)}-\Delta_{+}^{(N)}+\Delta_{-}^{(N)})/4.
\end{align} 
Hence, as well as an overall average level shift $\tilde{\delta}$, $\Delta_{x,y}$ couples in-plane modes to out-of-plane modes while preserving symmetry, $\Delta_{a,s}$ couples symmetric modes to antisymmetric modes while preserving direction, and $\Delta_{a,s;x,y}$ couples the symmetric in-plane to antisymmetric out-of-plane and vice-versa. For example, when a photon pulse is stored
in the out-of-plane modes and the shifts between the $m=\pm1$ levels are turned off, $\Delta_{a,s}$ is the only non-zero off-diagonal element in Eq.~(\ref{eq:dh}), coupling $\Pc_{xs} $ and $\Pc_{xa} $. This can be utilized in the control of the superposition of the stored photon and, hence, the emission direction of the photon upon retrieval, as described in Sec.~\ref{sec:storage}. 

Going beyond the regime $\lambda\lesssim a\ll \sqrt{A}$ and considering $a<\lambda$ is straightforward: The effective parameters described in Eqs.~(\ref{eq:td1}-\ref{eq:td2}) lead to identical coupling between modes, as described in Eq.~(\ref{eq:dh}), with $\Lambda$ found by numerically diagonalizing the full $3N\times 3N$ matrix $\mathcal{H}$ in Eq.~(\ref{eq:eom}). We then choose the eigenmodes which most closely resemble the ideal uniform symmetric and antisymmetric in-plane and out-of-plane modes.


%

\end{document}